\documentclass[aps,prd,twocolumn,showpacs,superscriptaddress,nofootinbib,preprintnumbers]{revtex4-2}

\usepackage{amsmath}
\usepackage{amsfonts}
\usepackage{amssymb}
\usepackage{enumerate}
\usepackage{color}
\usepackage{graphicx}
\usepackage{bm}
\usepackage[colorlinks=true,citecolor=blue,urlcolor=blue]{hyperref}
\usepackage{afterpage}
\usepackage{multirow}
\usepackage[table]{xcolor}

\newcommand{\Eq}[1]{Eq.~\eqref{#1}}

\begin{document}
\preprint{FERMILAB-PUB-21-613-T}
\preprint{UTTG-01-2022}

\title{Investigation of CMB constraints for dark matter-helium scattering}

\author{Kimberly K.~Boddy}
\affiliation{Theory Group, Department of Physics, The University of Texas at Austin, Austin, TX 78712}

\author{Gordan Krnjaic}
\affiliation{Theoretical Physics Department, Fermi National Accelerator Laboratory, Batavia, IL 60510}
\affiliation{Department of Astronomy and Astrophysics, University of Chicago,, Chicago, IL 60637}
\affiliation{Kavli Institute for Cosmological Physics, University of Chicago, Chicago, IL 60637}

\author{Stacie Moltner}
\affiliation{Theory Group, Department of Physics, The University of Texas at Austin, Austin, TX 78712}

\begin{abstract}
We study dark matter-helium scattering in the early Universe and its impact on constraints from cosmic microwave background (CMB) anisotropy measurements. We describe possible theoretical frameworks for dark matter-nucleon interactions via a scalar, pseudoscalar, or vector mediator; such interactions give rise to hydrogen and helium scattering, with cross sections that have a power-law dependence on relative velocity. Within these frameworks, we consider three scenarios: dark matter coupling to only neutrons, to only protons, and to neutrons and protons with equal strength. For these various cases, we use \textit{Planck} 2018 temperature, polarization, and lensing anisotropy data to place constraints on dark matter scattering with hydrogen and/or helium for dark matter masses between 10 keV and 1 TeV. For any model that permits both helium and hydrogen scattering with a non-negative power-law velocity dependence, we find that helium scattering dominates the constraint for dark matter masses well above the proton mass. Furthermore, we place the first CMB constraints on dark matter that scatters dominantly/exclusively with helium in the early Universe.
\end{abstract}

\maketitle

\section{Introduction}

Cosmic microwave background (CMB) data provide some of the best evidence for the existence of dark matter (DM)~\cite{Planck:2018vyg}.
The anisotropy of the CMB is well described by the standard $\Lambda$CDM cosmology, in which DM is a cold and collisionless matter component of the Universe.
However, many efforts to incorporate DM into the Standard Model (SM) of particle physics introduce interactions between DM and SM particles.
The early Universe offers a pristine environment to probe potential nongravitational scattering between dark and visible matter, without the astrophysical uncertainties that affect the interpretation of direct and indirect detection searches.

Elastic scattering between DM and visible matter induces a drag force between the DM and the baryon-photon fluids in the early Universe~\cite{Chen:2002yh}.
Such interactions damp perturbations on small scales, which can produce observable modifications to the CMB anisotropy power spectra.
The scattering cross section typically needs to be large---depending on the model---to have a measurable impact on the CMB, but a broad range of DM masses can be probed.

Previous studies have constrained scattering processes between DM and baryons under a variety of different assumptions, but they all incorporate scattering with at least hydrogen nuclei (i.e., protons)~\cite{Chen:2002yh,Sigurdson:2004zp,Dvorkin:2013cea,Gluscevic:2017ywp,Boddy:2018kfv,Xu:2018efh,Slatyer:2018aqg,Boddy:2018wzy}, aside from recent work that also has separate analyses for DM-electron scattering~\cite{Nguyen:2021cnb,Buen-Abad:2021mvc}.
DM scattering with protons immediately implies scattering with helium, unless the interaction is spin-dependent.
Incorporating helium scattering into CMB analyses generically improves constraining power, particularly for DM mass above 1~GeV~\cite{Gluscevic:2017ywp,Boddy:2018kfv,Xu:2018efh}.
However, the relationship between the scattering cross sections for helium and for hydrogen is model-dependent.
In the case of velocity-independent scattering, scattering may occur coherently on all nucleons in helium~\cite{Chen:2002yh} or on only the protons in helium~\cite{Chen:2002yh,Gluscevic:2017ywp,Boddy:2018kfv,Xu:2018efh}.
References~\cite{Boddy:2018kfv,Sigurdson:2004zp} also explored scattering on the protons in helium for velocity-dependent interactions.

Despite this rich variety of analyses, there has never been a dedicated CMB study of DM-\textit{neutron} scattering; indeed, the neutron is the only known particle present during recombination whose DM interactions have not been studied comprehensively, as DM-photon~\cite{Boehm:2001hm,Wilkinson:2013kia,Stadler:2018jin} and DM-neutrino~\cite{Wilkinson:2014ksa,Olivares-DelCampo:2017feq} scattering scenarios have been previously explored.
While Ref.~\cite{Chen:2002yh} did include the effects of DM-neutron interactions through scattering on helium, the analysis assumed coherent scattering on all nucleons in helium and did not account for the mass difference between hydrogen and helium in writing the DM-helium cross section; furthermore, the analysis was limited to velocity-independent scattering for DM mass $\gtrsim 1~\mathrm{GeV}$.
More generally, DM may scatter with both protons and neutrons with an arbitrary ratio of couplings.

In this paper, we use \textit{Planck} 2018 temperature, polarization, and lensing anisotropy measurements~\cite{Planck:2019nip} to obtain CMB constraints for various combinations of DM-neutron and DM-proton interactions.
We assume all neutrons are contained within $^4$He nuclei after big bang nucleosynthesis, neglecting the small abundances of other elements and isotopes.
Thus, our analysis involves DM scattering with hydrogen and helium only in the early Universe.
The scenarios of interest are as follows:
\begin{enumerate}[A]
\item \textbf{DM-neutron only:} DM scatters on neutrons bound in helium, and the interaction must be spin-independent.
\item \textbf{DM-proton only:} DM may interact with both hydrogen and helium.
  For spin-dependent (SD) interactions, DM scatters only with hydrogen, since helium has spin zero.
  Spin-independent (SI) interactions permit scattering with both hydrogen and helium.
\item \textbf{Equal proton/neutron couplings:} DM may interact with both hydrogen and helium (for SI interactions) or only hydrogen (for SD interactions).
  For SD interactions, this scenario is the same as Scenario B, and we refer to this case as Scenario B/C-SD.
\end{enumerate}

We discuss the motivation for models of DM-nucleon scattering and show how generic considerations give rise to cross sections that have power-law scalings of the relative particle velocity $v$.
Thus, our analyses parametrize the cross section as a power-law in $v$, and we use the generic models to set the relationship between the DM-helium and DM-hydrogen cross sections.
For our benchmark scenarios, we obtain constraints on the momentum-transfer cross section for DM masses ranging from 10~keV to 1~TeV and for velocity power-law indices $n \in \{-4, -2, 0, 2, 4\}$.

Our CMB analysis for Scenario A, in which DM scatters with neutrons and not protons, is the first to consider scattering with only helium.
Such DM-neutron interactions have been motivated by new-physics models attempting to explain the neutron lifetime anomaly~\cite{Fornal:2018eol} and the Atomki beryllium decay anomaly~\cite{Seto:2016pks}.
Additionally, various isospin-violating DM models have been motivated by direct detection anomalies in previous years~\cite{Feng:2011vu,Gao:2011bq,Yaguna:2016bga,Kelso:2017gib}.
While our analysis does not assume any specific scenario, the limits we derive are sufficiently general to constrain any of these scenarios for appropriately sized interactions strengths.

Throughout this paper, we use ``hydrogen'' and ``helium'' to reference the DM scattering target, while ``proton'' and ``neutron'' refer to the fundamental particle interaction (i.e., proton scattering can result in both hydrogen and helium scattering).
Additionally, when referring to the cross section, we always mean the momentum-transfer cross section,\footnote{The momentum-transfer cross section is obtained by weighting the differential cross section by the fractional longitudinal momentum transfer and integrating over all angles: $\sigma_T = \int d\Omega\, (1-\cos\theta) \frac{d\sigma}{d\Omega}$.} which is relevant for cosmology.
We often omit ``momentum-transfer'' for brevity.

In Section~\ref{sec:theory}, we discuss possible theoretical frameworks in which DM interacts with neutrons and protons.
In Section~\ref{sec:cosmology}, we treat DM interactions with baryons in a cosmological setting and present the corresponding modified Boltzmann equations.
We describe our analysis in Section~\ref{sec:analysis} and present our results in Section~\ref{sec:results}.
We conclude in Section~\ref{sec:conclusions}.

\section{Theory}
\label{sec:theory}

There are many viable ways to generate scattering interactions between dark and visible matter, and each scenario requires the addition of at least one new ``mediator" particle that connects the DM to quarks.
In this section, we survey some representative models that realize interactions with varying degrees of neutron-philic couplings and velocity/spin dependence.
These interactions induce DM-neutron scattering during the early Universe and therefore affect CMB anisotropies.

In Table~\ref{tab:models}, we present general results for the nonrelativistic DM-nucleus momentum-transfer cross section [using the notation of \Eq{eq:sigma} in Section~\ref{sec:cosmology}] for a generic set of Lorentz structures.
These formulas can be directly compared with the mediator and model choices in the following subsections.
In our notation convention, $c_i^{(\prime)}$ is a (pseudo) scalar coupling and $g_i$ is a vector coupling to particle species $i$.

\subsection{Scalar and pseudoscalar mediators}
\label{sec:scal}

If the mediating particle is a spin-0 scalar or pseudoscalar with renormalizable interactions, the most general Lagrangian contains the following terms:
\begin{equation}
  \mathcal{L}_\textrm{int} =
  \phi \, \bar{\chi} (c_\chi + i c_\chi^\prime \gamma^5) \chi + \phi \, \sum_q \bar{q} (c_q + i c^\prime_q \gamma^5) q \, ,
  \label{eq:leff}
\end{equation}
where $c_\chi^{(\prime)}$ is a (pseudo)scalar coupling to DM, and $c_q^{(\prime)}$ is a (pseudo)scalar coupling to SM quarks $q$.
We take the DM $\chi$ to be a Dirac fermion for simplicity.
In principle, these couplings in \Eq{eq:leff} are free parameters.
Following the conventions in Ref.~\cite{DelNobile:2013sia}, the induced $\phi$-nucleon $N$ coupling from this interaction can be written using the nuclear matrix element
\begin{equation}
\label{cN}
  \sum_q \langle N | c_q \bar{q} q | N \rangle \equiv c_N \bar{N} N \quad \text{(scalar)}
\end{equation}
where the relationship between the quark couplings ($c_q$) and nucleon couplings ($c_N$) is presented in the Appendix.

This discussion gives the most general parametrization of scalar-nucleon interactions, assuming either scalar or pseudoscalar couplings to quarks in the UV theory.
Here, we have exploited the freedom to choose arbitrary flavor structure without worrying about experimental constraints, which can be quite severe depending on the scenario.


\begin{table*}[ht]
    \centering
    \begin{tabular}{|c|c|c|c|c|c|}
    \hline
        \multirow{2}{*}{~ $n$ ~}
        & \multirow{2}{*}{\shortstack[c]{Interaction \\ $(\chi-N)$}}
        & \multirow{2}{*}{SI/SD}
        & Scenario A
        & Scenarios B, C
        & Scenarios B-SI ($\mathcal{C}=1$), C-SI ($\mathcal{C}=4$) \\
    \cline{4-6}
    & & & He scattering & H scattering & He scattering \\
    \hline \vspace{-8pt}&&&&&\\
    -4 & V-V, light & SI
    & $\displaystyle 2\pi  \frac{ g_\chi^2 g_N^2}{4\pi^2} \frac{2 \ln \left(2/\theta_c\right)}{\mu_{\chi \mathrm{He}}^2 }$
    & $\displaystyle 2\pi  \frac{ g_\chi^2 g_N^2}{4\pi^2} \frac{2 \ln \left(2/\theta_c\right)}{\mu_{\chi \mathrm{H}}^2 }$
    & $\displaystyle 4\mathcal{C} \left(\frac{\mu_{\chi \mathrm{H}}}{\mu_{\chi \mathrm{He}}} \right)^2 \sigma_{0, \mathrm{H}}$
    \\ \vspace{-8pt}&&&&&\\
    & S-S, light & SI
    & $\displaystyle 2\pi \frac{c_\chi^2 c_N^2}{16 \pi^2} \frac{2 \ln \left(2/\theta_c\right)}{\mu_{\chi \mathrm{He}}^2}$
    & $\displaystyle 2\pi \frac{c_\chi^2 c_N^2}{16 \pi^2} \frac{2 \ln \left(2/\theta_c\right)}{\mu_{\chi \mathrm{H}}^2}$
    & $\displaystyle 4\mathcal{C} \left(\frac{\mu_{\chi \mathrm{H}}}{\mu_{\chi \mathrm{He}}} \right)^2 \sigma_{0, \mathrm{H}}$
    \\ \noalign{\vspace{-8pt}} &&&&&\\
    \hline \noalign{\vspace{-8pt}} &&&&&\\
    -2 & S-P, light & SD
    & 0
    & $\displaystyle 4\pi \frac{c_{\chi}^2 c_N^{\prime 2}}{8\pi^2} \frac{1}{m_\mathrm{H}^2}$
    & 0
    \\ \noalign{\vspace{-8pt}} &&&&&\\
    & P-S, light & SI
    & $\displaystyle 4\pi \frac{c_{\chi}^{\prime 2} c_N^2}{32\pi^2} \frac{1}{m_\chi^2}$
    & $\displaystyle 4\pi \frac{c_{\chi}^{\prime 2} c_N^2}{32\pi^2} \frac{1}{m_\chi^2}$
    & $\displaystyle 4\mathcal{C} \sigma_{0, \mathrm{H}}$
    \\ \noalign{\vspace{-8pt}} &&&&&\\
    \hline \noalign{\vspace{-8pt}} &&&&&\\
    0 & V-V, heavy & SI
    & $\displaystyle 4\pi \frac{ g_\chi^2 g_N^2}{4\pi^2 m_V^4} \mu_{\chi \mathrm{He}}^2$
    & $\displaystyle 4\pi \frac{ g_\chi^2 g_N^2}{4\pi^2 m_V^4} \mu_{\chi \mathrm{H}}^2$
    & $\displaystyle 4\mathcal{C} \left(\frac{\mu_{\chi \mathrm{He}}}{\mu_{\chi \mathrm{H}}} \right)^2 \sigma_{0, \mathrm{H}}$
    \\ \noalign{\vspace{-8pt}} &&&&&\\
    & S-S, heavy & SI
    & $\displaystyle 4\pi \frac{c_\chi^2 c_N^2}{4 \pi^2 m_\phi^4}\mu_{\chi \mathrm{He}}^2$
    & $\displaystyle 4\pi \frac{c_\chi^2 c_N^2}{4 \pi^2 m_\phi^4} \mu_{\chi \mathrm{H}}^2$
    & $\displaystyle 4\mathcal{C} \left(\frac{\mu_{\chi \mathrm{He}}}{\mu_{\chi \mathrm{H}}} \right)^2 \sigma_{0, \mathrm{H}}$
    \\ \noalign{\vspace{-8pt}} &&&&&\\
    & P-P, light & SD
    & 0
    & $\displaystyle 4\pi \frac{ c_{\chi}^{\prime 2} c_N^{\prime 2}}{64\pi^2} \frac{\mu_{\chi \mathrm{H}}^2}{m_\chi^2 m_\mathrm{H}^2}$
    & 0
    \\ \noalign{\vspace{-8pt}} &&&&&\\
    \hline \noalign{\vspace{-8pt}} &&&&&\\
    2 & S-P, heavy & SD
    & 0
    & $\displaystyle \frac{16\pi}{3} \frac{c_{\chi}^2 c_N^{\prime 2}}{8\pi^2 m_\phi^4} \frac{\mu_{\chi \mathrm{H}}^4}{m_\mathrm{H}^2}$
    & 0
    \\ \noalign{\vspace{-8pt}} &&&&&\\
    & P-S, heavy & SI
    & $\displaystyle \frac{16\pi}{3} \frac{c_{\chi}^{\prime 2} c_N^2}{32\pi^2 m_\phi^4} \frac{\mu_{\chi \mathrm{He}}^4}{m_\chi^2}$
    & $\displaystyle \frac{16\pi}{3} \frac{c_{\chi}^{\prime 2} c_N^2}{32\pi^2 m_\phi^4} \frac{\mu_{\chi \mathrm{H}}^4}{m_\chi^2}$
    & $\displaystyle 4\mathcal{C} \left(\frac{\mu_{\chi \mathrm{He}}}{\mu_{\chi \mathrm{H}}} \right)^4 \sigma_{0, \mathrm{H}}$
    \\ \noalign{\vspace{-8pt}} &&&&&\\
    \hline \noalign{\vspace{-8pt}} &&&&&\\
    4 & P-P, heavy & SD
    & 0
    & $\displaystyle 8\pi \frac{ c_{\chi}^{\prime 2} c_N^{\prime 2}}{64\pi^2 m_\phi^4} \frac{\mu_{\chi \mathrm{H}}^6}{m_\chi^2 m_\mathrm{H}^2}$
    & 0
    \\ \noalign{\vspace{-8pt}} &&&&&\\
    \hline
    \end{tabular}
    \caption{Momentum-transfer cross section coefficients $\sigma_{0, B}$ for models with vector (V), scalar (S), and pseudoscalar (P) mediators.
    The first three columns list the power-law index $n$ for the velocity dependence of the cross section, the structure of the DM-nucleon interaction, and the dependence of the cross section on the nucleus spin (SI interactions permit DM-hydrogen and DM-helium scattering, while SD interactions permit DM-hydrogen scattering only).
    The remaining columns show expressions for $\sigma_{0, B}$ for helium scattering in Scenario A, hydrogen scattering in Scenarios B and C (relevant for both SI and SD interactions), and helium scattering in Scenarios B-SI and C-SI.
    For $n=-4$, $\sigma_{0, B}$ has a logarithmic divergence that we regulate with small cutoff angle $\theta_c$, determined by the details of a particular model.}
    \label{tab:models}
\end{table*}


\subsection{Vector mediators}
\label{sec:vec}

For spin-1 vector mediators, the coupling patterns to different quark flavors is constrained by the requirement that triangle Feynman diagrams cancel when the new gauge boson is an external leg of a 3-point diagram with virtual SM quarks (or other specified fields) flowing through the internal loop (see Ref.~\cite{Bauer:2018onh} for a discussion).
Models in which this cancellation occurs are anomaly-free and preserve unitarity; triangle diagram interactions grow with energy and eventually violate unitarity and, therefore, also spoil renormalizability.

DM candidates with masses well above the electroweak scale ($\gg$ 100 GeV), can be
charged under the weak force and interact with visible particles through the virtual exchange of known heavy particles (e.g., $W^\pm, Z^0$, or $h$).
By contrast, light ($\ll$ 100 GeV) DM with SM gauge charges would have been produced directly at collider experiments, which observed no new particles \cite{Egana-Ugrinovic:2018roi,Erler:2019hds}.
Since we cannot charge light DM under the SM gauge group, any model whose SM couplings do not automatically cancel triangle diagrams must feature additional (typically heavy) field content with appropriate SM charge assignments to restore this cancellation, which occurs automatically in the minimal SM with known field content.

\subsubsection{Minimal anomaly-free models}
\label{eq:anomalyfree}
There is a finite list of new abelian vectors that can be added without introducing anomalies.
Each such mediator $V$ corresponds to a SM interaction of the form
\begin{align}
  \mathcal{L}_\textrm{int} &= V_\mu J_\textrm{SM}^\mu \, , &
  J_\textrm{SM}^\mu &\equiv g \sum_f Q_f \bar{f} \gamma^\mu f \, ,
  \label{eq:Lvec}
\end{align}
where $g$ is an overall gauge coupling and the values of the $Q_f$ charges are given by anomaly cancellation requirements---for a review, see Ref.~\cite{Bauer:2018onh}.
For convenience, we define the overall coupling for species $f$ as $g_f \equiv g Q_f$.

The only anomaly free options without additional SM-charged fermionic field content are
\begin{equation}
  U(1)_{B-L}\, ,\quad
  U(1)_{B-3L_i}\, ,\quad
  U(1)_{L_i-L_j}\, ,
\end{equation}
where $B/L$ are baryon/lepton number and $L_i$ is a lepton family number.
The gauged $L_i - L_j$ scenario does not feature any couplings to quarks at tree level, so we ignore this possibility.
Note that in each of these cases, there is also an irreducible contribution to the $V-\gamma$ kinetic mixing parameter $\epsilon$, induced by loops of SM fermions with $V$ and $\gamma$ external legs.
This mixing in turn induces a $\epsilon V_\mu J_\textrm{EM}^\mu$ coupling to the SM electromagnetic current, but here $\epsilon \sim 10^{-2} g$ is generically suppressed.

For the $B-L$ and $B-3L_i$, the dependence on baryon number implies that $Q_f = 1/3$ for all quarks, which generates equal couplings at energies below the QCD confinement scale.
Thus,
\begin{equation}
  \mathcal{L} \supset \frac{g}{3} V_\mu \left(\bar{u} \gamma^\mu u + \bar{d} \gamma^\mu d \right)
  \to g V_\mu (\bar{p} \gamma^\mu p + \bar{n} \gamma^\mu n) \, ,
\end{equation}
and these scenarios can be constrained by both proton and neutron scattering in the early Universe.

Since SM anomaly cancellation need not affect the charge assignments for DM particles, we are free to choose the DM  coupling $g_\chi = g Q_\chi$ with an arbitrary value of $Q_\chi$ as long as the full particle content in the dark sector does not introduce additional, noncanceling triangle diagrams.

\subsubsection{Minimal ``anomalous'' models}
\label{sec:anomal}

Since no model is allowed to be anomalous without violating unitary/renormalizability, it is possible to patch anomalies by adding additional SM charged fields to cancel off the new triangle diagrams induced by an anomalous pattern of $U(1)$ charge assignments.
This enables vectors to couple to arbitrary currents of SM fields as long as viable ``anomalons'' can be added to cancel the corresponding triangle diagrams.

A popular example of this scenario is gauged $U(1)_B$~\cite{Dobrescu:2014ita}, which is phenomenologically similar to the $B-L$ example above, except there are no couplings to leptons; additional states are added instead to cancel anomalies, but these states can be sufficiently heavy that we can integrate them out well above our energy scales of interest.
For our purposes, the $U(1)_B$ model predicts equal proton/neutron couplings.
However, in principle, models in this category can be engineered to have arbitrary proton/neutron vector currents.
This class of scenarios is classified as ``minimal'' only to the extent that there is a single Abelian gauge group, even though other new fields are necessary to cancel anomalies; similar considerations apply to any arbitrary pattern of quark/lepton couplings.

\subsubsection{Beryllium-motivated nonminimal models}

The longstanding $\sim 7\sigma$ Atomki beryllium anomaly concerns a reported excess of events observed in the ${}^8\mathrm{Be}(18.15) \to {}^8\mathrm{Be}\, e^+e^-$ de-excitation, which may constitute evidence of a new $\approx$ 17 MeV particle~\cite{Krasznahorkay:2015iga} coupled to baryons and electrons.%
\footnote{However, see Ref.~\cite{Aleksejevs:2021zjw} for a recent interpretation involving only SM hadronic physics.}
Such light new particles must evade numerous experimental bounds.
It has been shown that viable models must violate isospin and couple preferentially to neutrons over protons \cite{Feng:2016jff}.

A leading candidate model to resolve the Atomki anomaly features the Lagrangian ~\cite{Feng:2016jff,Feng:2016ysn}
\begin{equation}
  \mathcal{L} = \frac{1}{4} X_{\mu\nu} X^{\mu\nu} + \frac{m_X^2}{2} X_\mu X^\mu - X_\mu J_{\rm SM}^\mu \, ,
  \label{eq:Latomki}
\end{equation}
where $X$ is a new vector boson and the SM current can be written as
\begin{equation}
  J^\mu_{\rm SM} \equiv  \sum_f g_f \bar{f} \gamma^\mu f \, ,
\end{equation}
where $f$ is a SM fermion.
Writing the SM-mediator coupling in units of the electric charge, $g_f \equiv e \epsilon_f$, the proton and neutron couplings are
\begin{align}
  \epsilon_p &= 2 \epsilon_u + \epsilon_d \, , &
  \epsilon_n &= \epsilon_u + 2 \epsilon_d~,
\end{align}
and to resolve the Atomki anomaly, the $X$ boson couplings must satisfy
\begin{equation}
  |\epsilon_p +\epsilon_n| \approx 0.011 \implies
  |\epsilon_u +\epsilon_d| \approx 3.7 \times 10^{-3} \, ,
  \end{equation}
and evading constraints from rare pion decay searches requires~\cite{Raggi:2015noa,Feng:2016ysn}
\begin{equation}
  |2 \epsilon_u + \epsilon_d| < 8 \times 10^{-4} \, .
\end{equation}
Thus, satisfying all of these requirements implies the relationship
\begin{equation}
  -0.067 < \frac{\epsilon_p}{\epsilon_n} < 0.078 \, ,
\end{equation}
so the proton coupling is sharply suppressed relative to the neutron coupling.

In addition to addressing the Atomki anomaly, the $X$ mediator can also consistently couple to DM if additional interactions are included.
For example, a Dirac DM particle $\chi$ can interact with $X$ if \Eq{eq:Latomki} is extended to include the operator
\begin{equation}
  \Delta {\cal L} = g_\chi X_\mu \bar \chi \gamma^\mu \chi \, ,
\end{equation}
which induces DM-nucleon scattering during the CMB era.

\subsection{Higher dimension operators}

Beyond the simple renormalizable interactions enumerated
above, it is possible to engineer a tower of operators with non-negative powers of momentum dependence of the form
\begin{equation}
  \mathcal{L}_\textrm{int} = \frac{1}{\Lambda^2} (\bar{\chi} \Gamma \chi) (\bar{f} \Gamma^\prime f) \, ,
\end{equation}
where $f$ is any SM fermion and $\Lambda$ is a new physics scale associated with the mass of a heavy particle, integrated out to yield this interaction.
The quantities $\Gamma$ and $\Gamma^\prime$ are each a linear combination of the following Lorentz structures:
\begin{equation}
  \gamma^\mu, \gamma^5,\gamma^\mu \gamma^5, q_\mu \sigma^{\mu \nu} \, .
\end{equation}
For most choices of $\Gamma$ and $\Gamma^\prime$, the corresponding cross section scales as $\propto v^{n}$, where $n$ can be realized using the renormalizable interactions from Sections~\ref{sec:scal} and \ref{sec:vec}.
However, for $n > 2$, the interaction must arise from a higher-dimension operator and goes beyond the above discussion.
For the remainder of this paper, we consider $n \in \{-4,-2,0,2,4\}$, and the $n=4$ case can only arise from a higher-dimension operator.

\section{Cosmology}
\label{sec:cosmology}

Within the theoretical framework of Section~\ref{sec:theory}, we can calculate the scattering quantity relevant for cosmology: the momentum-transfer cross section.
Relevant expressions are given in Table~\ref{tab:models} for the particular couplings to neutrons and protons in our Scenarios A, B, and C.
The cross sections all scale as power laws of the relative particle velocity $v$ with power-law index $n$.
Therefore, we parametrize the momentum-transfer cross section as
\begin{equation}
  \sigma_B (v) \equiv \sigma_{0, B} v^n
  \label{eq:sigma}
\end{equation}
where $B \in \{\mathrm{H}, \mathrm{He}\}$ denotes the particular particle species (i.e., hydrogen or helium) that DM scatters within the baryon fluid and $\sigma_{0,B}$ is a constant prefactor that our CMB analysis constrains.
In Scenarios B-SI and C-SI, in which there is scattering with hydrogen and helium, both cross sections scale with the same velocity dependence $n$.

Incorporating DM-baryon scattering in the early Universe requires modifying the standard Boltzmann equations~\cite{Ma:1995ey} that describe the evolution of perturbations.
We label quantities related to the DM and baryon fluids by $\chi$ and $b$, respectively.
In synchronous gauge, the time evolution of the density fluctuations $\delta_\chi$, $\delta_b$ and velocity divergences $\theta_\chi$, $\theta_b$ become~\cite{Chen:2002yh,Dvorkin:2013cea,Gluscevic:2017ywp,Boddy:2018kfv,Ma:1995ey}
\begin{equation}
  \begin{gathered}
     \dot{\delta}_\chi = -\theta_\chi - \frac{\dot{h}}{2} \, , \quad
    \dot{\delta}_b = -\theta_b - \frac{\dot{h}}{2} \, , \\
    \dot{\theta}_\chi = -\frac{\dot{a}}{a}\theta_\chi + c_\chi^2 k^2 \delta_\chi + R_{\chi} (\theta_b - \theta_\chi) \, , \\
    \dot{\theta}_b = -\frac{\dot{a}}{a}\theta_b + c_b^2 k^2 \delta_b + R_\gamma(\theta_\gamma - \theta_b) + \frac{\rho_\chi}{\rho_b} R_{\chi} (\theta_\chi - \theta_b) \, ,
    \end{gathered}
    \label{eq:Boltzmann}
\end{equation}
where $h$ is the trace of the scalar metric perturbation, $a$ is the scale factor, $k$ is the wave number, $c_\chi$ and $c_b$ are the sound speeds in each fluid, $\rho_\chi$ and $\rho_b$ are the energy densities, and overdots denote conformal time derivatives.
$R_\gamma$ and $R_{\chi}$ are the Compton scattering rate coefficient and the DM-baryon momentum-transfer rate coefficient, respectively.

The total momentum-transfer rate coefficient $R_\chi$ for DM scattering with hydrogen and helium is given by
\begin{equation}
  R_\chi = R_{\chi \mathrm{H}} + R_{\chi \mathrm{He}} \, ,
  \label{eq:rate}
\end{equation}
where
\begin{equation}
  R_{\chi B} = a \rho_b \frac{Y_B \sigma_{0,B} \mathcal{N}_n} {m_\chi + m_B}  \left( \frac{T_b}{m_B} + \frac{T_\chi}{m_\chi} \right)^{(1+n)/2}
  \label{eq:rateB}
\end{equation}
for each scattering species $B$~\cite{Boddy:2018kfv}.
In this expression, $Y_B$ is the mass fraction of species $B$, $\mathcal{N}_n \equiv 2^{(5+n)/2} \Gamma(3+n/2)/(3\sqrt{\pi})$, $m_\chi$ and $m_B$ are the DM and $B$ particle masses, and $T_\chi$ and $T_b$ are the DM and baryon fluid temperatures, respectively.

The coupled temperature evolution of the DM and baryon fluids is given by
\begin{equation}
  \begin{gathered}
    \dot{T}_\chi = -2 \frac{\dot{a}}{a} T_\chi + 2R'_{\chi} (T_b - T_\chi)\, , \\
    \dot{T}_b = -2 \frac{\dot{a}}{a} T_b + \frac{2\mu_b}{m_\chi} \frac{\rho_\chi}{\rho_b} R'_{\chi} (T_\chi - T_b) + \frac{2\mu_b}{m_e} R_\gamma (T_\gamma - T_b) \, ,
  \end{gathered}
\end{equation}
where $\mu_b \approx m_p (n_p + 4n_{\mathrm{He}})/(n_p + n_{\mathrm{He}} + n_e)$; $m_e$ is the mass of the electron; $m_p$ is the mass of the proton; and $n_e$, $n_p$, and $n_\textrm{He}$ are the number densities of electrons, protons, and helium, respectively.
The heat-exchange rate coefficient is given by
\begin{equation}
  R'_{\chi} = \frac{m_\chi}{m_\chi + m_\mathrm{H}} R_{\chi \mathrm{H}} + \frac{m_\chi}{m_\chi + m_\mathrm{He}} R_{\chi \mathrm{He}} \, .
  \label{eq:heatrate}
\end{equation}
In this work, we solve for the DM temperature, ignoring the backreaction on the baryon temperature evolution.
This approximate treatment is valid while the baryon and photon temperatures are tightly coupled, down to redshift $z\sim 300$.
Incorporating backreaction is expected to have little effect on our CMB analysis~\cite{Boddy:2018wzy}.

The velocities of the DM and baryon fluids are strongly coupled at early times for $n\geq 0$, rendering the relative bulk velocity of the fluids small, compared to the relative thermal velocities $v_\mathrm{th} = (T_b/m_B + T_\chi/m_\chi)^{1/2}$.
Negligible bulk velocities allows the velocity dependence of the cross section in \Eq{eq:sigma} to be governed by $v_\mathrm{th}$, as seen in \Eq{eq:rateB}.
For $n = -4$ and $n = -2$, however, interactions are suppressed at early times and the relative bulk velocity is significant around the time of recombination.
This complication introduces nonlinearities into the Boltzmann equations~\cite{Dvorkin:2013cea,Boddy:2018wzy}, and approximate methods can be employed to maintain linearity when numerically solving.
We follow Refs.~\cite{Dvorkin:2013cea,Xu:2018efh,Slatyer:2018aqg} by modifying the momentum-transfer rate coefficient in \Eq{eq:rateB} to be
\begin{equation}
  R_{\chi B} = a \rho_b \frac{Y_B \sigma_{0,B} \mathcal{N}_n} {m_\chi + m_B} \left(\frac{T_b}{m_B} + \frac{T_\chi}{m_\chi} + \frac{V_{\mathrm{RMS}}^2}{3}\right)^{(1+n)/2} \, ,
  \label{eq:rate-vrms}
\end{equation}
where $V_{\mathrm{RMS}}$ is the root mean square (RMS) relative bulk velocity between the DM and baryon fluids.
Under $\Lambda$CDM, the RMS velocity is given by $cV_\textrm{RMS} \sim 30~\mathrm{km/s}$ prior to recombination ($z \gtrsim 1000$) and scales as $(1+z)^2$ at smaller redshifts.
The $\Lambda$CDM evolution for $V_\textrm{RMS}$ is applicable for our CMB analysis with 100\% of DM interacting; more advanced techniques must be used to analyze scenarios in which only a fraction DM interacts~\cite{Boddy:2018wzy}.

\section{Analysis}
\label{sec:analysis}

We use \textit{Planck} 2018 data to constrain DM interactions, under our three scenarios of interest: Scenario A in which DM interacts only with neutrons, Scenario B in which DM interacts only with protons, and Scenario C in which DM interacts with both neutrons and protons with equal coupling strength.
Since $^4$He is a spin-0 nucleus, Scenario A only gives 
rise to SI scattering during the CMB era; Scenarios B and C may have either SI or SD interactions, corresponding to scattering with hydrogen and helium or with hydrogen only.
Our analysis uses a modified version\footnote{\url{https://github.com/kboddy/class_public/tree/dmeff}} of the Cosmic Linear Anisotropy Solving System (\texttt{CLASS})~\cite{Blas_2011} to solve the Boltzmann equations with the modifications described in Section~\ref{sec:cosmology} that incorporate DM scattering.

We sample our parameter space using the \texttt{cobaya} Bayesian analysis framework~\cite{Torrado:2020dgo,2019ascl.soft10019T} with the Markov chain Monte Carlo (MCMC) sampler~\cite{Lewis:2002ah,Lewis:2013hha} and fast-dragging~\cite{2005math......2099N}.
We use the \textit{Planck} 2018 likelihood code and employ the \texttt{commander} and \texttt{simall} likelihoods for low multipoles, the \texttt{Plik lite} nuisance-marginalized joint likelihood for high multipoles, and the \texttt{SMICA} lensing reconstruction likelihood~\cite{Planck:2019nip}.

For each velocity power law $n$ and each neutron/proton coupling scenario, we sample the DM-baryon cross section $\sigma_{0, B}$ as a free parameter with a flat prior for seven fixed DM masses.
We also sample the following five standard $\Lambda$CDM cosmological parameters with broad flat priors: the Hubble parameter $H_0$, baryon density $\Omega_b h^2$, scalar amplitude $A_s$, scalar spectral index $n_s$, reionization optical depth $\tau$, and DM density $\Omega_\chi h^2$.
We assume all DM is interacting.

For Scenario A, DM scatters only with helium, so the sampling parameter is $\sigma_{0,\mathrm{He}}$.
For Scenario B/C-SD, DM scatters only with hydrogen, so the sampling parameter is $\sigma_{0,\mathrm{H}}$.
Scenarios B-SI and C-SI involve scattering on both hydrogen and helium; for these cases, we sample the parameter $\sigma_{0,\mathrm{H}}$ and fix the helium cross section according to its relation to the hydrogen cross section in Table~\ref{tab:models}.

Our analysis covers DM masses from 10 keV to 1 TeV.
Below 10 keV, the validity of our assumption of thermalized, cold DM breaks down for $n\geq 0$.%
\footnote{The DM temperature is below the baryon-photon temperature at early times for $n<0$~\cite{Boddy:2018wzy}, potentially allowing our analysis to be extended to lower DM masses.}
For DM masses much larger than the masses of hydrogen and helium, the DM parameters $\sigma_0$ and $m_\chi$ are degenerate, appearing together as $\sigma_{0,B}/m_\chi$ in the expression for $R_\chi$.
Our exclusion limits in Section~\ref{sec:results}, including scenarios that involve both hydrogen and helium scattering, scale as $\sigma_{0, B} \propto m_\chi$ at large DM masses.
Thus, our limits at $m_\chi = 1~\mathrm{TeV}$ can be extrapolated to larger DM masses.

\begin{figure*}[t]
  \centering
  \includegraphics[width=0.48\textwidth]{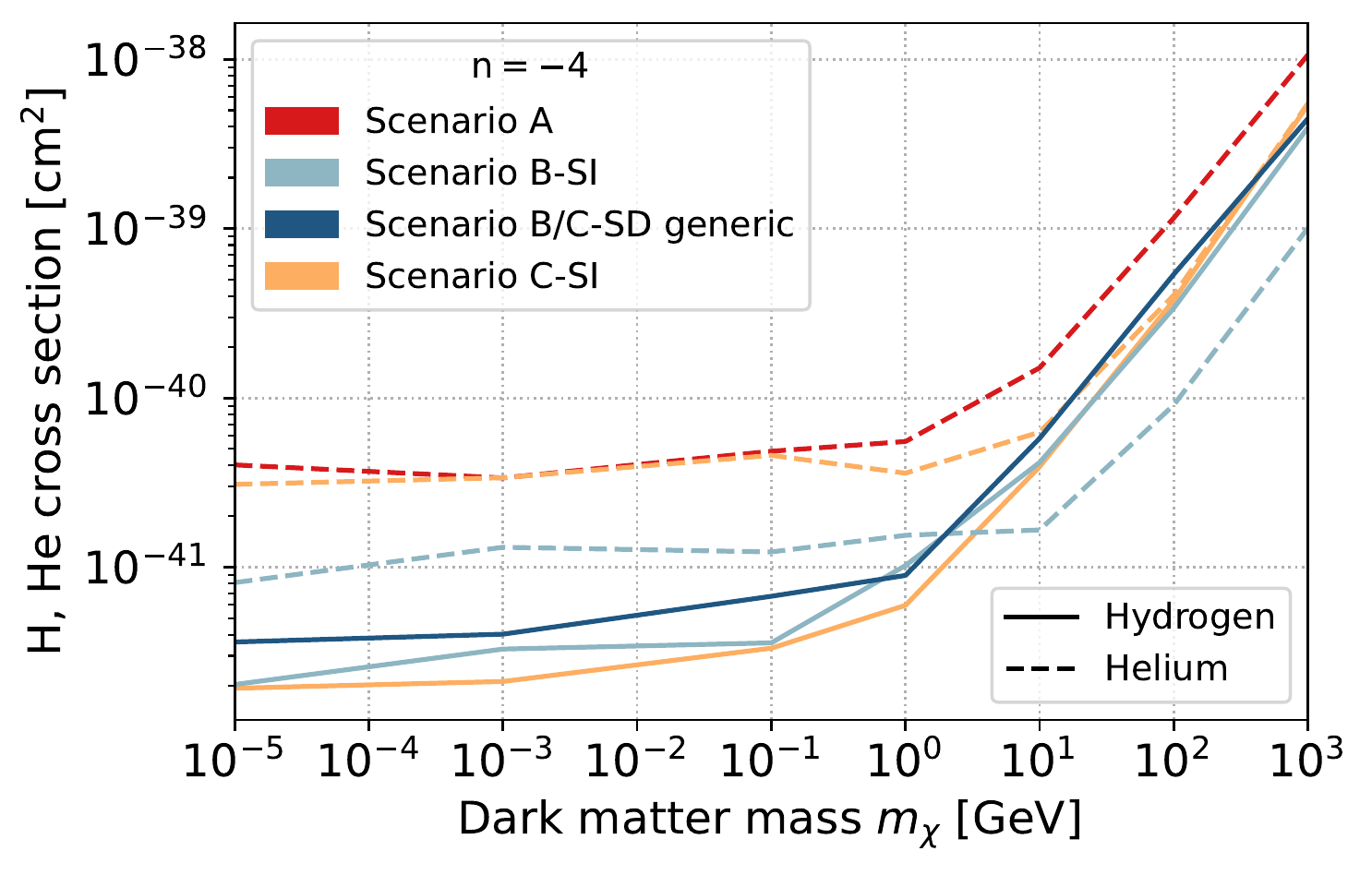}
  \includegraphics[width=0.48\textwidth]{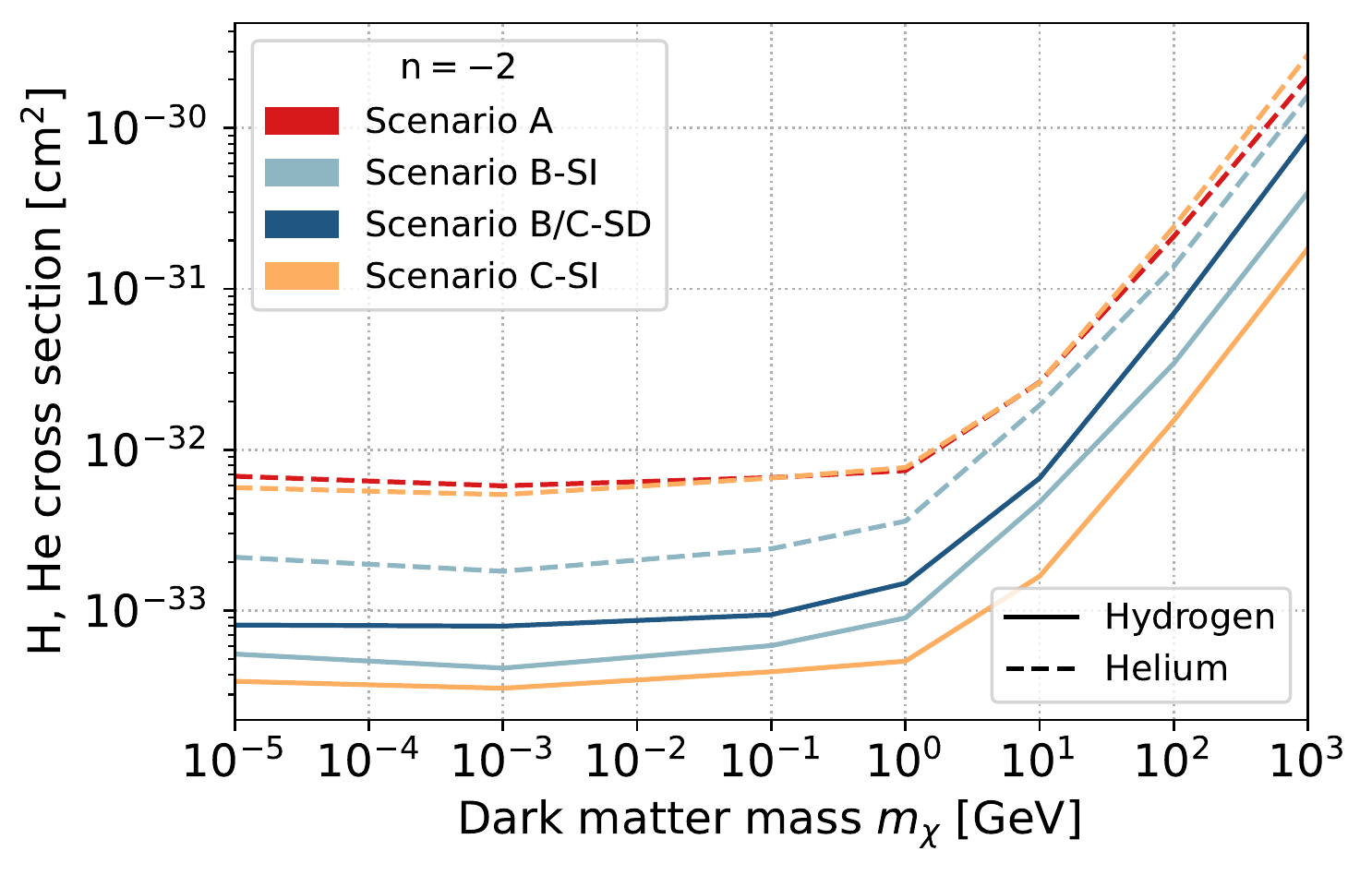} \\
  \includegraphics[width=0.48\textwidth]{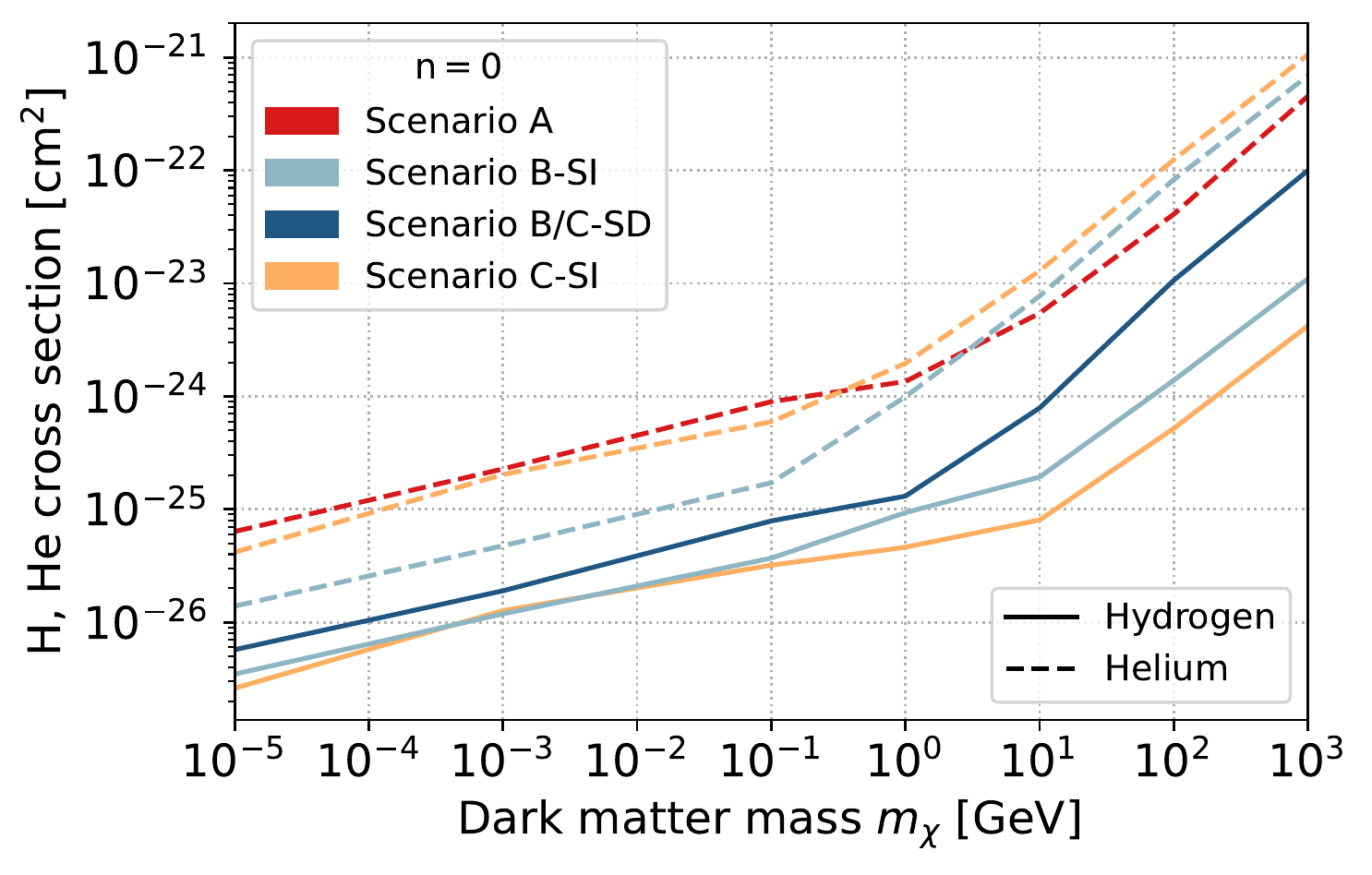}
  \includegraphics[width=0.48\textwidth]{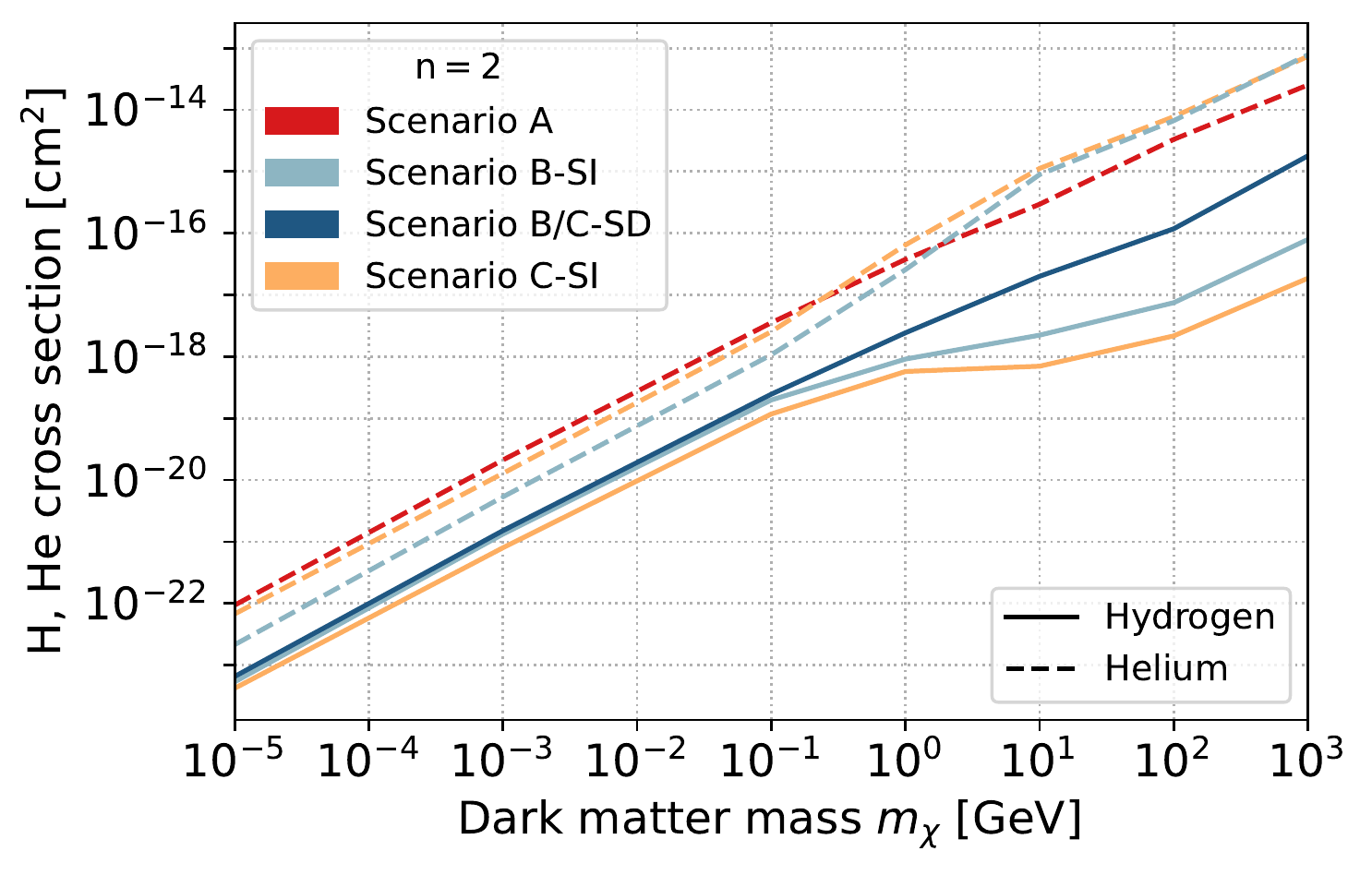} \\
  \includegraphics[width=0.48\textwidth]{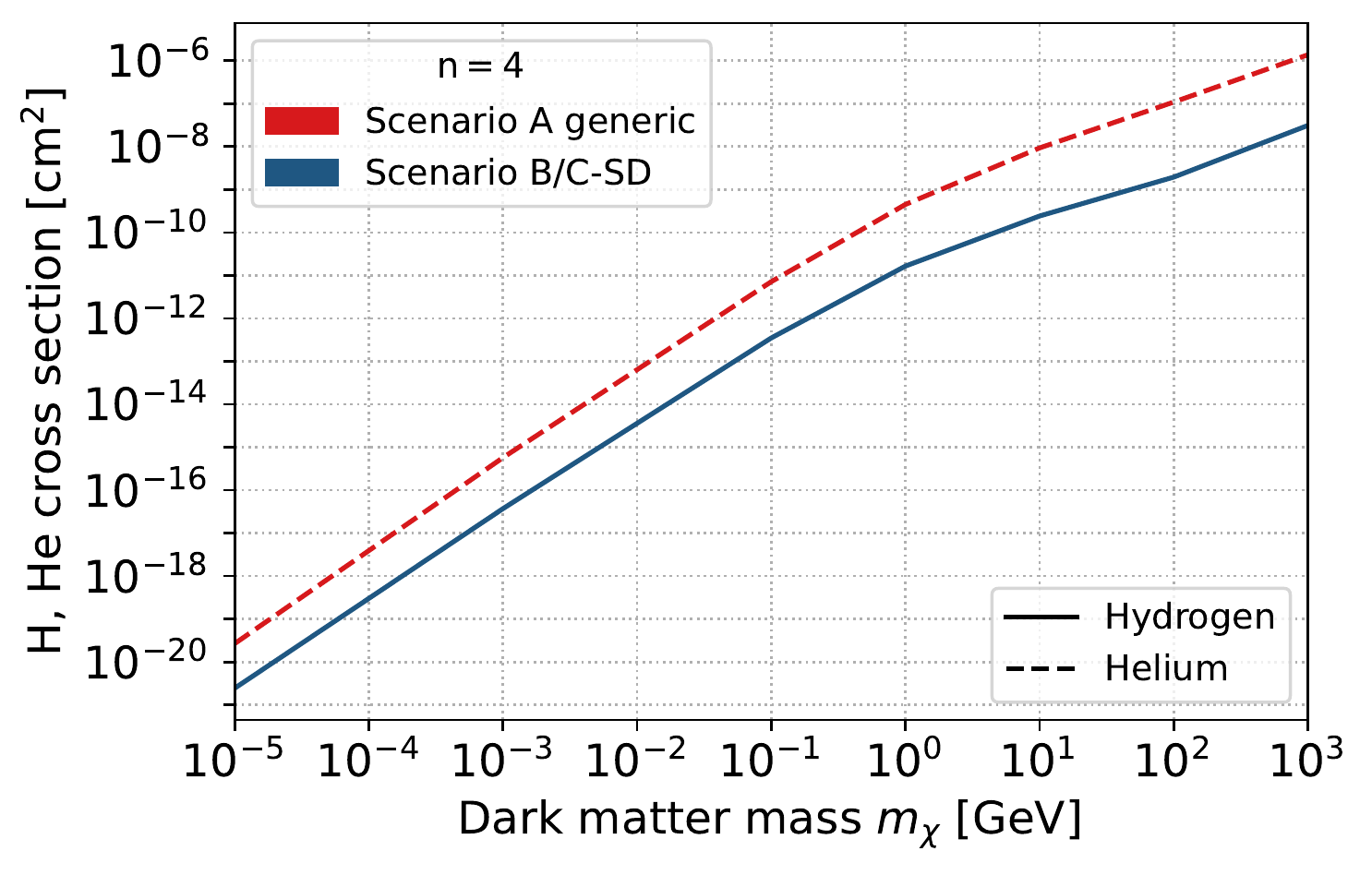}
  \caption{The 95\% C.L.\ upper limits on the momentum-transfer cross section coefficient $\sigma_{0,B}$, defined in \Eq{eq:sigma}, for DM-hydrogen (solid) and DM-helium (dashed) scattering.
    We show results for cross sections with a power-law velocity dependence of $n=-4$ (upper left), $n=-2$ (upper right), $n=0$ (center left), $n=2$ (center right), and $n=4$ (bottom).
    For each value of $n$, we analyze Scenario A with only a DM-neutron coupling (red), Scenario B-SI with only a DM-proton coupling (light blue), Scenario C-SI with equal DM couplings to protons and neutrons (orange), and Scenario B/C-SD (dark blue) for SD scattering.
    For Scenarios B-SI and C-SI, the hydrogen and helium cross sections are related by the expressions given in Table~\ref{tab:models}.
    The cases labeled as ``generic'' do not arise in the models presented in Section~\ref{sec:theory}, but we include them for completeness.}
  \label{fig:npow}
\end{figure*}

\section{Results}
\label{sec:results}

We present the 95\% confidence level (C.L.) upper limits on the momentum-transfer cross section between DM and hydrogen/helium as a function of DM mass for various velocity dependencies in Fig.~\ref{fig:npow}.
These results are also provided in Table~\ref{tab:limits} and as supplementary text files for convenience.

Scenario B/C-SD, corresponding to DM scattering with only hydrogen, has been studied previously, and we have verified consistency with recent work~\cite{Nguyen:2021cnb}.
For purposes of comparison, we include results for hydrogen-only scattering for $n=-4$, even though there is not an associated model in Table~\ref{tab:models}.
We label this case as ``generic'' in Fig.~\ref{fig:npow}.
We also include ``generic'' results for $n=4$ for Scenario A, corresponding to helium-only scattering.
Since this work places the first bounds on DM that preferentially scatters with helium rather than hydrogen, we include the result for $n=4$ so that there are limits on helium-only scattering for all values of $n$ in this study.

The impact of helium scattering depends on its contribution to the total momentum-transfer rate coefficient.
For Scenarios B-SI and C-SI, we can write $R_\chi = R_{\chi\mathrm{H}} (1 + R_{\chi\mathrm{He}}/R_{\chi\mathrm{H}})$, where ratio $R_{\chi\mathrm{He}}/R_{\chi\mathrm{H}}$ dictates the relative importance of helium scattering.
This ratio incorporates the model-dependent ratio of the cross sections.
From Table~\ref{tab:models}, we have 
\begin{equation}
  \frac{ \sigma_{0,\mathrm{He}} }{  \sigma_{0,\mathrm{H}} } = 4 \mathcal{C}  \left(\frac{  \mu_{\chi \mathrm{He}}   }{  \mu_{\chi \mathrm{H}}   } \right)^{n+2} \, ,
\end{equation}
where $\mathcal{C}=1$ for Scenario B-SI and $\mathcal{C}=4$ for Scenario C-SI.
For each of these scenarios, we determine $R_{\chi\mathrm{He}}/R_{\chi\mathrm{H}}$ analytically in the limits of large ($m_\chi \gg m_\mathrm{H}, m_\mathrm{He}$) and small ($m_\chi \ll m_\mathrm{H}, m_\mathrm{He}$) DM masses.
In our estimates below, we use a helium mass fraction $Y_\mathrm{He} \simeq 0.24$ and a ratio of masses $m_\mathrm{He}/m_\mathrm{H} \simeq 4$.

In the limit of large DM mass, the ratio of the momentum-transfer rates for Scenario B-SI or C-SI is
\begin{equation}
  \frac{R_{\chi \mathrm{He}}}{R_{\chi \mathrm{H}}} \approx 4\mathcal{C}\frac{Y_\mathrm{He}}{Y_\mathrm{H}}\left(\frac{m_\mathrm{He}}{m_\mathrm{H}}\right)^{n+2}\times
  \begin{cases}
    \left(\frac{m_\mathrm{H}}{m_\mathrm{He}}\right)^{(1+n)/2} & \!\!\! n\geq 0 \\
    1 &  \!\!\! n < 0 \, .
  \end{cases}
  \label{eq:ratehigh}
\end{equation}
For $n<0$ the RMS relative bulk velocity is larger than the thermal relative velocity at the redshift of interest~\cite{Boddy:2018wzy}, so here we have assumed the $V_\mathrm{RMS}$ term dominates and thus cancels upon taking the ratio of the rates.
Numerically, we have $R_{\chi \mathrm{He}} / R_{\chi \mathrm{H}} \simeq 10.1 \mathcal{C} \times 2^n$ for $n \geq 0$ and $R_{\chi \mathrm{He}} / R_{\chi \mathrm{H}} \simeq 20.2 \mathcal{C} \times 4^n$ for $n<0$.
Therefore, we expect the inclusion of helium scattering to be negligible for $n=-4$, comparable to hydrogen scattering for $n=-2$, and dominant for $n\geq 0$.
This behavior is evident in our numerical results for the extreme cases of $n=-4$ and $n=2$.
For $n=-4$, we find that Scenarios B-SI and C-SI have very similar bounds on $\sigma_{0,\mathrm{H}}$ as Scenario B/C-SD, indicating that the incorporation of helium scattering in Scenarios B-SI and C-SI has little impact on the resulting bound and thus hydrogen scattering drives the constraint.
In contrast, for $n=2$, the limits for Scenarios B-SI and C-SI have similar bounds on $\sigma_{0,\mathrm{He}}$ as Scenario A (helium-only scattering), indicating helium scattering drives the constraint.

In the limit of small DM mass, the ratio of the momentum-transfer rate coefficients is
\begin{equation}
  \frac{R_{\chi \mathrm{He}}}{R_{\chi \mathrm{H}}} \approx 4\mathcal{C}\frac{Y_\mathrm{He}}{Y_\mathrm{H}} \frac{m_\mathrm{H}}{m_\mathrm{He}} \simeq 0.32\mathcal{C} \, ,
  \label{eq:ratelow}
\end{equation}
so the contribution to the rate from helium scattering is subdominant to that from hydrogen scattering for Scenario B-SI and comparable for Scenario C-SI.
We, therefore, expect limits on $\sigma_{0,\mathrm{H}}$ for Scenarios B-SI and B/C-SD to coincide, as demonstrated by our results in Fig.~\ref{fig:npow} for $n=2$ in particular.
We also note that our limits on $\sigma_{0,\mathrm{He}}$ for Scenarios A and C-SI are close at low DM mass for the various $n$, but the limit for Scenario C-SI is slightly stronger, as both hydrogen and helium scattering contribute to the constraint.
Moreover, the limit on $\sigma_{0,\mathrm{He}}$ for Scenario B-SI is noticeably different, since hydrogen scattering is expected to drive the constraint.

We emphasize the relation between the helium and hydrogen cross sections within Scenario B-SI or C-SI is determined by the model, as given in Table~\ref{tab:models}.
For $n=-4$ with large DM masses, the helium cross section is smaller than (for Scenario B-SI) or equal to (for Scenario C-SI) the hydrogen cross section; otherwise, $\sigma_{0,\mathrm{He}} > \sigma_{0,\mathrm{H}}$.
Our results in Fig.~\ref{fig:npow} reflect these relations by construction.
In particular, a limit on $\sigma_{0,\mathrm{H}}$ that is lower than the corresponding $\sigma_{0,\mathrm{He}}$ of a given scenario does not indicate the data are more sensitive to hydrogen scattering.
On the contrary, we have found that helium scattering is the dominant effect in constraining DM interactions for large DM masses for $n \geq 0$.

\section{Conclusions}
\label{sec:conclusions}

In this paper, we conduct the first in-depth investigation of DM-helium scattering in the early Universe.
We account for the appropriate form and velocity dependence of the hydrogen and helium momentum-transfer cross sections that arise from heavy and light scalar, pseudoscalar, and vector mediators.
The cross sections exhibit a power-law dependence on relative velocity, with a power-law index $n \in \{-4, -2, 0, 2, 4\}$.
We also consider three scenarios for DM-nucleon couplings: DM-neutron only coupling, DM-proton only coupling, and equal coupling to protons and neutrons.
Using \textit{Planck} 2018 anisotropy data, we obtain the 95\% C.L. upper limits on the hydrogen and helium momentum-transfer cross sections for these different scenarios with different velocity dependencies.

Our results can be interpreted in the context of particular DM models, including those presented in Section~\ref{sec:theory}, to obtain limits on associated coupling constants and mediator masses.
However, since these limits constrain such large ($\sim$ barn sized) cross sections for models with $n\geq 0$ in Fig.~\ref{fig:npow}, the mediator-SM coupling must be fairly large and the mediator masses must be fairly light.
Thus, each model is also subject to additional laboratory constraints in the parameter space that realizes such cross sections (for examples, see Refs.~\cite{Tulin:2014tya,Batell:2014yra}, which show strong
constraints on the quark-mediator coupling at low mediator mass).
Although the constrained value of the mediator-SM coupling in a given model depends on the ratio of dark and visible couplings for a given cross section limit, perturbative unitarity for the DM coupling requires $c_{i} \lesssim 4\pi$ for all species \cite{PhysRevLett.64.615}; thus, for each choice of the dark/visible coupling ratio, there is a corresponding limit on $c_{N}$ that can realize the $\sigma_{0,B}$ we constrain in our analysis (see Table~\ref{tab:models}).

Given the model dependence of such coupling limits in each scenario, it is beyond the scope of our analysis to provide a direct comparison with the experimental limits in specific cases, but it is expected that for each of the constraint curves shown in Fig.~\ref{fig:npow} with $n \neq 0$, there are stronger laboratory bounds once the mediator mass and its SM couplings are specified within a given model (subject to unitarity bounds on the DM-mediator coupling).
Nonetheless, our results directly constrain the scattering properties of DM itself during the CMB era, without reference to any other hypotheses and, therefore, offer a new probe of protophobic interactions, particularly in the low ($<$ GeV) DM mass range where direct detection sensitivity thresholds are too high to probe the typical momentum transfers that DM in the halo imparts to nuclear targets.

We note that some of our constrained parameter space in Fig.~\ref{fig:npow} involves $\sigma_0 \gtrsim 10^{-25}~\mathrm{cm}^{2}$, where theoretical considerations invalidate the point-particle approximation for DM scattering with nucleons under the Born approximation~\cite{Digman:2019wdm,Cappiello:2020lbk}.
Larger cross sections can be achieved through enhancements via the exchange of multiple mediators or by considering composite DM states (e.g.\ dark nuclei~\cite{Krnjaic:2014xza}).
For the latter case, there are additional (model- and momentum-dependent) form factors that rescale the cross section with nucleons, which we do not include in our analysis here; thus, our results in which $\sigma_0 \gtrsim 10^{-25}~\mathrm{cm}^{2}$ are valid in the limit where these form factors are negligible for typical CMB era momentum transfers (e.g.\ when the momentum transfer is small compared to a given compositeness scale).
Since momentum transfers are set by the sub-eV photon temperature during the CMB era, we expect that any potential form factor suppression should be negligible throughout our parameter space of interest.
However, such form factors might be relevant for direct detection in the halo where momentum transfers can be larger; thus, comparing our limits to those of terrestrial scattering experiments (e.g. from Ref.~\cite{Monteiro:2020wcb}, which constrains neutron-philic composite DM with direct detection) might require a nontrivial mapping, depending on the nature of the form factor suppression. 

In comparing limits for point particle interactions, our results are complementary in mass range to existing bounds on neutron-philic DM from direct detection experiments.
For comparison, Ref.~\cite{Dey:2020sai} finds that GeV-scale DM with spin dependent DM-neutron cross section $\sigma_{\chi n} \approx 10^{-33}$ cm$^2$ can explain the XENON1T excess \cite{XENON:2020rca} through the Migdal effect while evading other direct detection bounds.
This model is constrained by our  limits on Scenario B/C-SD in the $n=0$ panel of Fig.~\ref{fig:npow}.
Near the $\sim$ GeV mass range, our limits ($\sigma_{\chi n} \lesssim 10^{-24}$ cm$^2$) are not sufficient to exclude the XENON1T preferred region in Ref.~\cite{Dey:2020sai}, but they extend the generic direct-detection limits on neutron-philic DM by many orders of magnitude toward lower mass where such limits were previously unavailable. 

Despite strong laboratory constraints on particular models, our CMB bounds provide valuable and complementary information on the interaction properties of cosmologically abundant particle DM.
Upcoming ground-based CMB experiments, such as the Simons Observatory~\cite{SimonsObservatory:2018koc} and CMB-S4~\cite{Abazajian:2019eic}, will achieve significant improvements in angular resolution, compared to existing data.
Therefore, advancements in CMB experiments will lead to better sensitivity to DM scattering physics, which suppresses structure more at smaller scales.

\begin{acknowledgments}
We thank Vera Gluscevic for useful discussions.
The work at UT is supported in part by the National Science Foundation (NSF) under Grant No.~PHY-2112884.
G.K.\ is supported by the Fermi Research Alliance, LLC under Contract No.~DE-AC02-07CH11359 with the U.S. Department of Energy, Office of Science, Office of High Energy Physics.
This work was performed in part at the Aspen Center for Physics, which is supported by NSF Grant No.~PHY-1607611.

The authors acknowledge the Texas Advanced Computing Center (TACC) at The University of Texas at Austin for providing high performance computing resources used to run the MCMC chains for our analysis.
\end{acknowledgments}

\begin{table*}[ht]
\centering
\begin{tabular}{|c|>{\centering}r@{\ }l|>{\centering}m{2.2cm}|>{\centering}m{2.2cm}|>{\centering}m{2.2cm}|>{\centering}m{2.2cm}|>{\centering}m{2.2cm}|>{\centering\arraybackslash}m{2.2cm}|}
\hline
\multirow{2}{1cm}{\centering $n$}
& \multicolumn{2}{c|}{\multirow{2}{1.4cm}{\centering DM Mass}}
& Scenario A
& \multicolumn{2}{c|}{Scenario B-SI}
& Scenario B/C-SD
& \multicolumn{2}{c|}{Scenario C-SI}\\
\cline{4-9}
& & & He scattering
& H scattering & He scattering
& H scattering
& H scattering & He scattering \\
\hline
-4
& 10 & keV & 4.0e-41 & 2.0e-42 & 8.1e-42 & 3.6e-42* & 1.9e-42 & 3.1e-41 \\
& 1 & MeV & 3.4e-41 & 3.3e-42 & 1.3e-41 & 4.0e-42* & 2.1e-42 & 3.4e-41 \\
& 100 & MeV & 4.8e-41 & 3.6e-42 & 1.2e-41 & 6.7e-42* & 3.3e-42 & 4.6e-41 \\
& 1 & GeV & 5.5e-41 & 1.0e-41 & 1.5e-41 & 9.0e-42* & 6.0e-42 & 3.6e-41 \\
& 10 & GeV & 1.5e-40 & 4.2e-41 & 1.7e-41 & 5.8e-41* & 3.9e-41 & 6.3e-41 \\
& 100 & GeV & 1.2e-39 & 3.4e-40 & 9.1e-41 & 5.4e-40* & 3.8e-40 & 4.0e-40 \\
& 1 & TeV & 1.1e-38 & 4.0e-39 & 1.0e-39 & 4.5e-39* & 5.4e-39 & 5.5e-39 \\
\hline
-2
& 10 & keV & 6.8e-33 & 5.4e-34 & 2.1e-33 & 8.1e-34 & 3.6e-34 & 5.8e-33 \\
& 1 & MeV & 6.0e-33 & 4.4e-34 & 1.8e-33 & 8.0e-34 & 3.3e-34 & 5.3e-33 \\
& 100 & MeV & 6.7e-33 & 6.0e-34 & 2.4e-33 & 9.4e-34 & 4.2e-34 & 6.7e-33 \\
& 1 & GeV & 7.4e-33 & 9.0e-34 & 3.6e-33 & 1.5e-33 & 4.8e-34 & 7.7e-33 \\
& 10 & GeV & 2.6e-32 & 4.7e-33 & 1.9e-32 & 6.7e-33 & 1.6e-33 & 2.6e-32 \\
& 100 & GeV & 2.1e-31 & 3.4e-32 & 1.4e-31 & 7.0e-32 & 1.5e-32 & 2.4e-31 \\
& 1 & TeV & 2.1e-30 & 4.0e-31 & 1.6e-30 & 9.0e-31 & 1.8e-31 & 2.9e-30 \\
\hline
0
& 10 & keV & 6.4e-26 & 3.5e-27 & 1.4e-26 & 5.7e-27 & 2.6e-27 & 4.2e-26 \\
& 1 & MeV & 2.3e-25 & 1.2e-26 & 4.8e-26 & 1.9e-26 & 1.3e-26 & 2.0e-25 \\
& 100 & MeV & 9.0e-25 & 3.7e-26 & 1.7e-25 & 7.9e-26 & 3.2e-26 & 6.0e-25 \\
& 1 & GeV & 1.4e-24 & 9.4e-26 & 9.9e-25 & 1.3e-25 & 4.6e-26 & 2.0e-24 \\
& 10 & GeV & 5.4e-24 & 1.9e-25 & 7.7e-24 & 7.9e-25 & 8.0e-26 & 1.3e-23 \\
& 100 & GeV & 4.1e-23 & 1.4e-24 & 8.3e-23 & 1.1e-23 & 5.2e-25 & 1.2e-22 \\
& 1 & TeV & 4.5e-22 & 1.1e-23 & 6.9e-22 & 1.0e-22 & 4.2e-24 & 1.1e-21 \\
\hline
2
& 10 & keV & 9.4e-23 & 5.4e-24 & 2.1e-23 & 6.6e-24 & 4.2e-24 & 6.8e-23 \\
& 1 & MeV & 2.1e-20 & 1.3e-21 & 5.3e-21 & 1.5e-21 & 8.0e-22 & 1.3e-20 \\
& 100 & MeV & 3.5e-18 & 2.0e-19 & 1.1e-18 & 2.5e-19 & 1.2e-19 & 2.5e-18 \\
& 1 & GeV & 3.8e-17 & 9.1e-19 & 2.6e-17 & 2.4e-18 & 5.7e-19 & 6.5e-17 \\
& 10 & GeV & 3.0e-16 & 2.2e-18 & 8.9e-16 & 2.0e-17 & 7.0e-19 & 1.1e-15 \\
& 100 & GeV & 3.3e-15 & 7.5e-18 & 6.7e-15 & 1.2e-16 & 2.2e-18 & 7.7e-15 \\
& 1 & TeV & 2.5e-14 & 7.9e-17 & 7.8e-14 & 1.8e-15 & 1.9e-17 & 7.3e-14 \\
\hline
4
& 10 & keV & 2.7e-20* &  &  & 2.4e-21 &  &  \\
& 1 & MeV & 5.7e-16* &  &  & 3.7e-17 &  &  \\
& 100 & MeV & 7.2e-12* &  &  & 3.5e-13 &  &  \\
& 1 & GeV & 4.5e-10* &  &  & 1.7e-11 &  &  \\
& 10 & GeV & 9.3e-09* &  &  & 2.4e-10 &  &  \\
& 100 & GeV & 1.1e-07* &  &  & 2.0e-09 &  &  \\
& 1 & TeV & 1.4e-06* &  &  & 3.1e-08 &  &  \\
\hline
\end{tabular}
\caption{
The 95\% C.L. upper limits on the momentum-transfer cross section coefficient $\sigma_{0,B}$, corresponding to the limits shown in Fig.~\ref{fig:npow}.
The cross sections have a power-law dependence on relative velocity, with a power-law index $n$.
Scenario A corresponds to DM coupled only to neutrons, Scenario B to DM coupled only to protons, and Scenario C to DM with equal coupling to neutrons and protons.
For Scenarios B and C, the interaction may be SI or SD; note the scenarios are equivalent for the SD interaction.
Blank entries do not have a corresponding model from Section~\ref{sec:theory}.
$^\ast$We include $n=-4$ results for Scenario B-SD and $n=4$ results for Scenario A, even though they are not represented by one of the models.}
\label{tab:limits}
\end{table*}


\appendix
\setcounter{secnumdepth}{0}
\section{Appendix: Scalar formalism}
\label{sec:formalism}

In this section, we review the formalism for defining the mediator-nucleon coupling at low energy in terms of mediator-quark interactions in the high energy theory (above the scale of QCD confinement).
Here we follow the conventions of Ref.~\cite{DelNobile:2013sia}.

The induced mediator-nucleon coupling in \Eq{cN} is
\begin{equation}
  c_N = \sum_{q = u,d,s} c_q \frac{m_N}{m_q} f_{Tq}^{(N)} + \frac{2}{27} f_{TG}^{(N)}
  \sum_{q = c,b,t} c_q \frac{m_N}{m_q} \, ,
\end{equation}
and we have defined the parameters
\begin{align}
  f^{(N)}_{Tq} &\equiv \langle N | \frac{m_q \bar{q} q}{m_N} | N \rangle \, , &
  f^{(N)}_{TG} &= 1 - \sum_{q=u,d,s} f^{(N)}_{Tq} \, ,
\end{align}
where $f_{Tq}^{(N)}$ represents light quark contributions to the nucleon mass and can be found in the appendix of Ref.~\cite{DelNobile:2013sia}.
Similarly, the pseudoscalar nucleon matrix element can be written
\begin{equation}
  \langle N | \sum_q c^\prime_q \bar{q} \gamma^5 q |N \rangle
  \equiv i c^\prime_N \bar{N} \gamma^5 N \quad \text{(pseudo)}
\end{equation}
where the effective nuclear coupling can be written
\begin{align}
  c_N^\prime &= \sum_{q=u,d,s} \frac{m_N}{m_q} \left(c_q -C \right) \Delta^{(N)}_q \, , &
  C &\equiv \bar{m} \sum_q \frac{c_q^\prime}{m_q} \, ,
\end{align}
and we have defined $\bar{m} \equiv (1/m_u + 1/m_d + 1/m_s)^{-1}$.
The parameters $\Delta^{(N)}_q$ satisfy the relation
\begin{equation}
  \Delta^{(N)}_q s^\mu \equiv \langle N | \bar{q} \gamma^\mu \gamma^5 q | N \rangle \, ,
\end{equation}
where the new parameters here can be found in the appendix of Ref.~\cite{DelNobile:2013sia}.

\bibliography{dmeff.bib}

\end{document}